\documentclass[reprint,twocolumn,amssymb]{revtex4}
\newcommand{\be}{\begin{equation}}
\newcommand{\ee}{\end{equation}}
\newcommand{\bea}{\begin{eqnarray}}
\newcommand{\eea}{\end{eqnarray}}
\usepackage{graphicx}
\usepackage{amsmath}
\usepackage{color}
\usepackage{dcolumn}% Align table columns on decimal point
\usepackage{bm}% bold math
\usepackage{hyperref}% add hypertext capabilities
\usepackage{braket}
\usepackage{enumerate}

\begin{document}

\title{Bulk density signatures of a lattice quasihole with very few particles}
\author{R. O. Umucal\i lar}
\affiliation{Department of Physics, Mimar Sinan Fine Arts University, 34380 Sisli, Istanbul, Turkey}

\date{\today}

\begin{abstract}
Motivated by the recent experimental realization of a two-particle fractional quantum Hall state of ultracold atoms in a small optical lattice [Nature 619, 495 (2023)], we propose a minimal setup to create and observe a quasihole in such a system. We find that clear signatures of a quasihole state with two or three atoms can be obtained through a standard site-resolved density measurement provided that the system is appropriately modified with simple additional potential profiles. By adding a single-site repulsive potential to pin the quasihole and superimposing a harmonic trap on top of the optical lattice to keep the particles away from the system edge, we determine via exact diagonalization an optimal range for system parameters such as the magnetic flux and the strengths of the additional potentials that would favour the creation of the quasihole state. We hope that our results will be a useful guide for a possible proof-of-principle experiment that will demonstrate the first controllable creation of a simple quasihole state in a condensed matter system, which will pave the way for the observation of the anyonic statistics of quasiholes in a more complex system.
\end{abstract}

\maketitle

%%%
{\it Introduction.---}
%%%
Starting with the discovery of the fractional quantum Hall (FQH) effect in a two-dimensional electron gas \cite{quantum Hall experiment}, the physics community has put great effort into observing the theorized fractional statistics of its quasiparticles \cite{fractional statistics} and a certain subgroup of these quasiparticles, the so-called Abelian anyons have recently been detected in electronic systems \cite{FQH experiments}. The more elusive type of non-Abelian anyons, which hold promise for topological quantum computation \cite{anyon computation review}, are yet to be discovered \cite{non-abelian anyons}. In order to achieve a greater control in the manipulation of these exotic quasiparticles, researchers have been searching for the same physics in different systems like the ultracold atomic or polaritonic ones, which have already been proved to be prolific platforms for quantum simulations \cite{quantum simulation}. 

The search for fractional quantum Hall physics of ultracold atoms, which started with the proposals to create an effective magnetic field for neutral atoms both in continuum and in an optical lattice\cite{synthetic field}, has culminated in the realization of this effective magnetic field \cite{Hofstadter experiments} and a recent observation of a two-particle FQH state \cite{Greiner2023}. The next logical steps seem to be the realization of an FQH state with a larger number of particles and the observation of the fractionally charged quasiparticles, and it is our aim in this article to propose a minimal setup for a possible proof-of-principle experiment to create and detect such a quasiparticle in an experimentally realistic small system.

Owing to certain advantages of optical lattices in the creation of the magnetic field and the enhancement of the energy gap above the ground state, numerous studies have been performed for the lattice, some using artificial periodic boundary conditions to study the bulk properties \cite{PBC lattice} and some with open (or hard-wall, box) boundaries to study the edge properties or rather to connect with realistic experiments \cite{box lattice}. In the case with open boundaries, the concept of filling fraction, that is the ratio of the number of particles to the number of magnetic flux quanta, which is a defining property of an FQH state, is not well defined especially in a small system. In this work, building on the ideas of our previous works \cite{Umucalilar2018b, Macaluso2020}, we show that the lattice system can adjust the filling factor properly by itself due to the competing effects of the incompressiblity of the correlated FQH-like states and the tendency of a superimposed harmonic potential to accumulate the particle cloud in the center. We suggest that by only measuring the site densities in the presence and absence of a repulsive potential localized at a lattice site to pin a quasihole, one can construct two quantities, namely, the ratio between the mean-square-radii of the clouds in the two cases and the density depletion created due to the pinning potential, the joint observation of which will be a clear-cut demonstration of the quasihole state even with two or three particles. We also provide phase diagrams for these quantities over a wide range of parameters, which we hope will be a useful guide for future experiments.

%%%
{\it The Model.---}
%%%
The starting point of our exact diagonalization study is the well-known Hofstadter-Bose-Hubbard Hamiltonian for bosonic particles in a tight-binding square lattice with complex hopping phases and on-site interactions, modified with the pinning and harmonic confinement potentials:
\begin{multline}
H = - t\sum_{\langle ij\rangle}\left(e^{i2\pi \phi_{ij}} 
c^\dag_{i}c_{j}+\text{h.c.}\right)+\frac{U}{2}\sum_i n_i(n_i-1)
\\+Vn_0+\Omega\sum_i r_i^2n_i, 
\label{eq:Hamiltonian}
\end{multline}
where $c^{\dagger}_i$ ($c_j$) creates (annihilates) a 
boson at site $i$ ($j$), $n_i = c^{\dagger}_i c_i$ is the number operator, \text{h.c.} is the Hermitian conjugate, and $t > 0$ is the hopping amplitude between nearest-neighbor sites $\langle ij\rangle$  with coordinates ${\bf r}_i$ and ${\bf r}_j$. We adopt the symmetric gauge ${\bf A} = (B/2)(x{\bf \hat{y}}-y{\bf \hat{x}})$ to make comparisons with the usual ansatz wave functions for the continuum and to determine the experimentally relevant hopping phase $\phi_{ij} = (1/\phi_0) \int_{{\bf r}_j}^{{\bf r}_i} {\bf A}\cdot d{\bf r}$, where the integration path is a straight line and $\phi_0 = h/q_0$ is the magnetic flux quantum for an effective charge $q_0$. This choice of the vector potential corresponds to an effective perpendicular magnetic field along the $z$ direction with strength $B$ and the magnetic flux quantum per unit cell of the square lattice with separation $a$ is defined as $\phi = Ba^2/\phi_0$. The wave function of a particle traversing a loop around the unit cell acquires the Aharonov-Bohm phase factor $\exp(i2\pi\phi)$. This result and the density signatures we propose are gauge invariant. However, if one uses a different gauge ${\bf A} + \nabla \Lambda({\bf r}) $ and wants to make a wave-function comparison, the ansatz many-particle wave functions must be multiplied by the factor $\exp[-i\sum_j\Lambda({\bf r}_j)/\phi_0]$.

The strength of the repulsive on-site interactions between particles is quantified by $U>0$ and that of the single-site pinning potential located at the central site ($i = 0$) is given by $V>0$. This central site is also chosen to be the origin of our coordinate system and the imposed harmonic potential with strength $\Omega>0$.

%%%
{\it Mean-square-radius and Density Depletion.---}
%%%
In two of our previous studies \cite{Umucalilar2018a,Macaluso2020}, we made use of the following relation between the mean-square-radius $\langle r^2\rangle$ of the particle cloud and the expected value of its total angular momentum $\langle L_z \rangle$ in order to extract information about the statistical phase due to the braiding of quasiholes: 
\bea
\langle r^2\rangle = \frac{2\ell_B^2}{N}\bigg(\frac{\langle L_z\rangle}{\hbar} + N\bigg),
\label{eq:r2 Lz}
\eea
where $\ell_B = \sqrt{\phi_0/2\pi B}$ is the magnetic length and $N$ is the number of particles in the system. This relation has also been made use of in later works to characterize certain quasiparticle properties like the statistical phase, charge, and spin for various FQH states including non-Abelian ones \cite{mean-square works}. In \cite{Macaluso2020}, we investigated a moderate-sized ($16\times 16$) lattice system with a relatively large number of particles ($N = 12, 18$) via a tensor-network method. We also exploited the stabilizing effect of an additional harmonic potential in obtaining the Laughlin-type states (as we do in the present work); however, we did not make a systematic study of the dependence of results on the magnetic field and harmonic potential strengths as such a study would be numerically very costly. Since we were focused on the braiding phase, we also overlooked the fact that an experimentally realistic system with edges containing as few particles as $N =2,3$ could still be interesting in its own right, especially as a showcase for a simple demonstration of the ansatz Laughlin quasihole. Here, we close that gap. We put forward two density signatures to observe Laughlin-type physics, by comparing two cases that differ only by the absence or presence of the pinning potential. We will see that in a wide parameter range: (i) the ratio between the mean-square-radii [Eq. (\ref{eq:r2 Lz})] in these two cases turns out to be a very good indicator of this physics especially when the number of particles is low \cite{Umucalilar2018b} and (ii) the density depletion caused by the introduction of the pinning potential is very close to the expected continuum result. Measurement of these two density-dependent observables together will be an excellent indication that the created states are indeed Laughlin-type states.

Let us briefly recall the ansatz wave functions Laughlin suggested for a microscopic explanation of the FQH effect \cite{Laughlin1983}:
\bea
\Psi_{\rm L}(\zeta_1, \ldots, \zeta_N) \propto \prod_{j<k}(\zeta_j-\zeta_k)^m e^{-\sum_{i = 1}^N|\zeta_i|^2/4\ell_B^2}\label{eq:WF_Laughlin},\\
\Psi_{\rm QH}(\{\zeta_i\},\mathcal{Q}) \propto  \prod_{i=1}^N(\zeta_i\!-\!\mathcal{Q})\Psi_{\rm L}(\zeta_1, \ldots, \zeta_N),
\label{eq:WF_QH}
\eea
where $\Psi_{\rm L}$ and $\Psi_{\rm QH}$ represent the Laughlin and one-quasihole states, respectively, up to normalization, $\zeta_j = x_j+iy_j$ is the complex-valued coordinate of the $j$th particle, $\mathcal{Q}$ is the complex-valued coordinate of the quasihole, and $\nu = 1/m$ is the Landau-level filling fraction. Laughlin originally used these wave functions to explain the $\nu = 1/3$ effect but the ansatz was shown in numerous studies to extend to other fractions $\nu = 1/m$, $m$ being an odd (even) integer for fermions (bosons). 

We will focus on the $m = 2$ case for bosonic atoms and set the quasihole coordinate to be $\mathcal{Q} = 0$, supposing that it is pinned at the origin, when it is pinned. For this configuration, the continuum Laughlin and one-quasihole states are both total angular momentum eigenstates with eigenvalues $N(N-1)\hbar$ and $[N(N-1)+N]\hbar = N^2\hbar$, respectively. If we insert these values in Eq. (\ref{eq:r2 Lz}), we get
\bea
\langle r^2\rangle_{\rm L} = 2N\ell_B^2, \:\:\: \langle r^2\rangle_{\rm QH} = 2(N+1)\ell_B^2.
\label{eq:r^2 QH}
\eea
The relation $\ell_B = \sqrt{\phi_0/2\pi B} = a/\sqrt{2\pi\phi}$ can be used to determine the lattice counterparts of these quantities; however, although for moderate $\phi$ ($\sim 0.15$) and small enough $\Omega a^2/t$ ($\sim
0.005$) the results are pretty close to the continuum ones, significant deviations develop outside of this regime \cite{Supplemental}. Instead, as we did in a previous work in which we considered a lattice with periodic boundary conditions \cite{Umucalilar2018b}, we propose to use the ratio between the quantities in Eq. (\ref{eq:r^2 QH}) as one of our indicators of the lattice ground states, which turns out to be a quite robust quantity:
\bea
\mathcal{R}_{\rm ideal} \equiv \langle r^2\rangle_{\rm L}/\langle r^2\rangle_{\rm QH} = N/(N+1).\label{eq:R L/QH}
\eea
The ratio in Eq. (\ref{eq:R L/QH}) compares $\langle r^2\rangle = \sum_i r_i^2\langle n_i \rangle/N$ for the lattice Laughlin and one-quasihole states with the same $N$, $\phi$, and $\Omega$; the only difference is that $\langle r^2\rangle_{\rm QH}$ is to be measured in the case where $V\neq 0$ so as to pin a quasihole. Our rationale is that if the system ground state is very different from the quasihole state, the ratio $\mathcal{R} \equiv \langle r^2\rangle_{V = 0}/\langle r^2\rangle_{V \neq 0}$ significantly deviates from $\mathcal{R}_{\rm ideal} $, as we confirmed numerically. Note that this value becomes indiscernible with growing particle number, and as such, $\mathcal{R}$ is well-suited only for the cases with small particle numbers that we are investigating. 

As for our second observable, the density depletion, we again compare the two cases with and without a pinning potential, other parameters being the same, and calculate how much density is displaced outwards in a given region with radius $r$ by the repulsive pinning potential due to the incompressiblity of the Laughlin state:
\bea 
(\Delta n)_{r} \equiv \sum_{r_i<r}[\langle n_i\rangle_{V = 0} -\langle n_i\rangle_{V \neq 0}].
\label{eq:DEP}
\eea
In a system with periodic boundary conditions, this quantity saturates at a certain value indicating the fractional density depletion of the quasihole \cite{Zeybek2022}, but in a finite-sized droplet it gradually rises to a maximum value and then gradually vanishes as $r$ grows to contain all particles. In our numerical simulations, we take this maximum value $\Delta n \equiv {\rm max}\{(\Delta n)_{r}\}$ as the density depletion. For a large enough system it can be proven that this quantity equals the filling fraction ($\nu = 1/2$), but for small number of particles there occur deviations from this value, which can be determined by numerical integration of the continuum densities.

\begin{figure}[ht]
\includegraphics[width=0.45\textwidth]{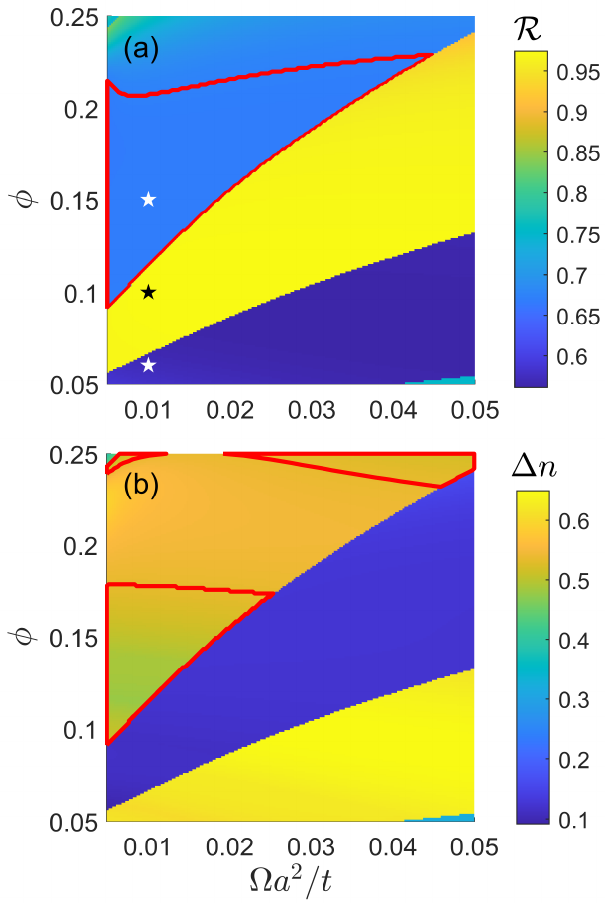}
\caption{Phase diagrams for $N=2$ in the plane of the harmonic trap strength ($\Omega$) and flux quanta per unit cell ($\phi$). (a) Ratio $\mathcal{R}$ between the mean-square-radii found for $V=0$ and $V\neq 0$. Stars at $\Omega a^2/t = 0.01$ and $\phi = 0.06, 0.10, 0.15$ indicate the parameters used for Fig. \ref{fig:Site occupations 2p}. (b) Maximum value $\Delta n$ of the depleted density due to the pinning potential, evaluated as in Eq. (\ref{eq:DEP}). Red lines surround regions with $0.660<\mathcal{R}<0.673$ in (a) and $0.475<\Delta n<0.535$ in (b).}
\label{fig:Phase diagrams}
\end{figure}

%%%
{\it Numerical Results.---}
%%%
In our numerical simulations for $N=2$, we considered an $11\times 11$ lattice with open boundaries. Fixing the interaction and pinning strengths at $U = 7t$ (which is close to the value in \cite{Greiner2023}) and $V = 10t$, we made an extensive parameter scan in the $\Omega-\phi$ plane and constructed `phase  diagrams' for our indicators $\mathcal{R}$ and $\Delta n$ [Fig. (\ref{fig:Phase diagrams})] (see \cite{Supplemental} for the effect of changing $V$). For the case of $N = 3$ particles in a $13\times13$ lattice, due to the large size of the Hilbert space, we assumed that the interactions are hardcore (at most one particle at a site) and tabulated the results for a small number of points in the $\Omega-\phi$ plane \cite{Supplemental}. 

\begin{figure}[ht]
\includegraphics[width=0.48\textwidth]{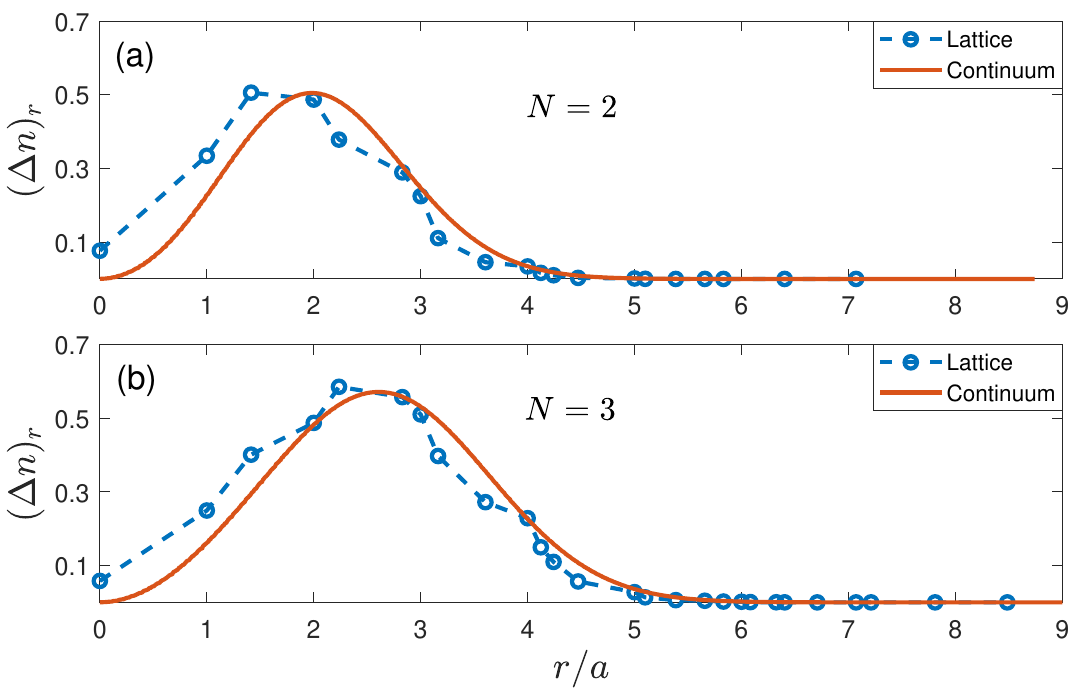}
\caption{Density depletion $(\Delta n)_r$ due to a pinned quasihole as a function of the distance $r$ from the pinning site for (a) $N=2$ ($\Omega = 0.01$, $\phi = 0.15$) and (b) $N=3$ ($\Omega = 0.007$, $\phi = 0.125$). Also shown by red lines are the numerically integrated results using the continuum wave functions $\Psi_{\rm L,QH}$. Maximum values of this integration are $\Delta n = 0.505,0.571$, for $N=2,3$, respectively.}
\label{fig:Depletions}
\end{figure}

In Fig. \ref{fig:Depletions}, we show two samples of $(\Delta_n)_r$ for $N = 2,3$ as a function of $r$ together with the numerically integrated continuum counterparts, for which the distance unit is converted to $a$ using its relation to $\ell_B$. For the chosen parameters lattice results agree well with the continuum ones, with apparent discrepancy for small $r$ as the region considered contains few lattice sites. As we discussed in the previous section, we are interested in the maximum value of $(\Delta_n)_r$ and take it as the missing density $\Delta n$ at the position of the quasihole, which is then used to generate Fig. \ref{fig:Phase diagrams}(b). The agreement for this maximum value between lattice and continuum results is quite well as can be observed from Fig. \ref{fig:Depletions}.

Turning to the phase diagrams, in Fig. \ref{fig:Phase diagrams}, it is remarkably seen that wide ranges of our indicators with nearly constant values are separated with quite sharp boundaries, for which we will offer an explanation shortly. Red lines in the figure delineate regions where the values of the indicators are within a certain percentage of the expected values, namely, one percent around $\mathcal{R}_{\rm ideal} = 2/(2+1) = 2/3$ ($N = 2$) and six percent around $(\Delta n)_{\rm ideal} = 0.505$ (found by numerical integration). Generally speaking, parameters in the roughly triangular upper left region of the $\Omega-\phi$ plane (whose lower corner starts from $\phi\sim 0.1$) seem to be quite conducive to the realization of the  quasihole state; however, $\Delta n$ seems to be more sensitive to parameter changes and a more truthful choice could be the smaller triangular region in panel (b) remaining inside the delineated region of panel (a). In this region the overlaps between the system ground states and the ansatz states are also very high ($\sim 99 \%$).

\begin{figure}[ht]
\includegraphics[width=0.48\textwidth]{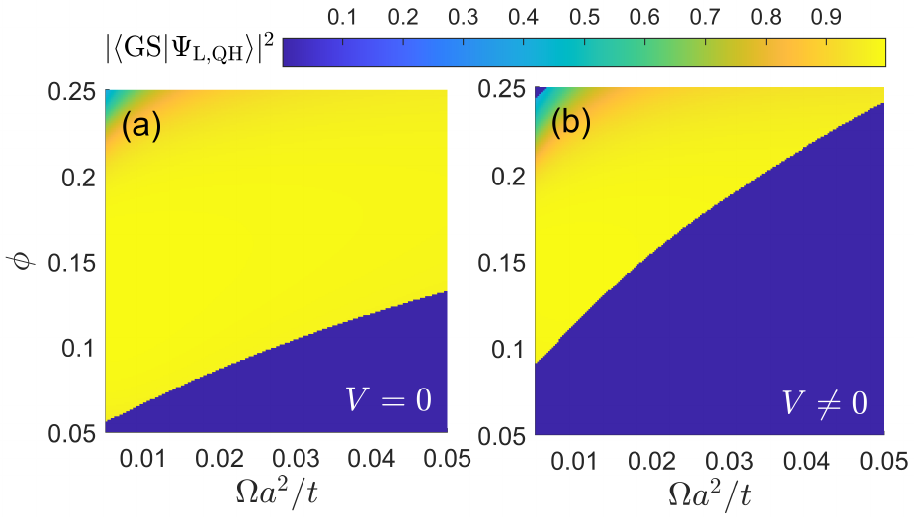}
\caption{Overlaps between the system ground states and the ansatz states represented by Eqs. (\ref{eq:WF_Laughlin}, \ref{eq:WF_QH}) for $N=2$ in the plane of the harmonic trap strength ($\Omega$) and flux quanta per unit cell ($\phi$) ($1.0$ corresponds to $100\%$ overlap).  (a) Laughlin-state overlap $\mathcal{O}_{\rm L} = |\langle {\rm GS}|\Psi_{\rm L}\rangle|^2$ for $V = 0$. (b) Quasihole-state overlap $\mathcal{O}_{\rm QH} = |\langle {\rm GS}|\Psi_{\rm QH}\rangle|^2$ for $V \neq 0$.}
\label{fig:Overlaps}
\end{figure}

In order to understand the sharp boundaries better and also to provide more evidence for the Laughlin-type physics, in Fig. \ref{fig:Overlaps} we show the overlaps $\mathcal{O}_{\rm L,QH} = |\langle {\rm GS}|\Psi_{\rm L,QH}\rangle|^2$ between the numerical ground states $|{\rm GS}\rangle$ and the corresponding continuum states projected onto the lattice given by $|\Psi_{\rm L,QH}\rangle \propto \sum \Psi_{\rm L,QH}(\zeta_1, \ldots, \zeta_N)c^{\dagger}_{\zeta_1} \ldots c^{\dagger}_{\zeta_N}|{\rm vac}\rangle $, where the sum is over all possible particle coordinates to be chosen at lattice sites. While Fig. \ref{fig:Overlaps}(a) displays the overlap with the Laughlin state in the absence of a pinning potential ($V = 0$), Fig. \ref{fig:Overlaps}(b) is for the overlap with the quasihole state when $V \neq 0$. Comparison of Figs. \ref{fig:Phase diagrams} and \ref{fig:Overlaps} clearly shows that the sharp boundaries in the phase diagrams correspond to sudden changes in the overlaps; the uppermost one is due to changes in the quasihole-state overlap and the lower one is caused by changes in the Laughlin-state overlap. We also checked that these sudden changes in the overlaps are accompanied by the closing and reopening of the energy gap above the ground state, which is characteristic of topological transitions \cite{Supplemental}. 

\begin{figure}[ht]
\includegraphics[width=0.48\textwidth]{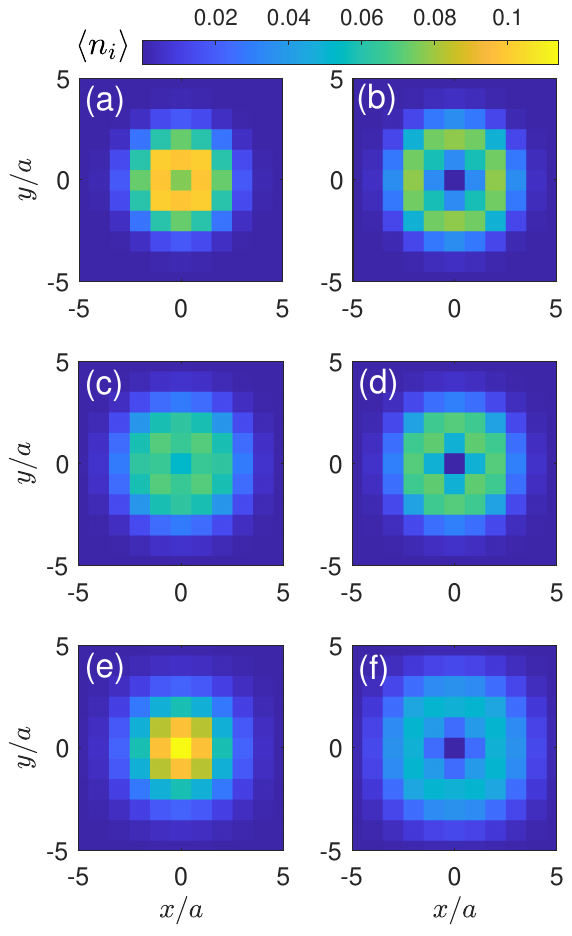}
\caption{Site occupations $\langle n_i\rangle$ for $N=2$ particles with fixed harmonic trap strength $\Omega a^2/t = 0.01$ and three different flux quanta per unit cell (a-b) $\phi = 0.15$, (c-d) $\phi = 0.1$, (e-f) $\phi = 0.06$, corresponding to stars in Fig. \ref{fig:Phase diagrams}(a). Left panels are for $V = 0$ and the right ones for $V \neq 0$.}
\label{fig:Site occupations 2p}
\end{figure}

It can also be observed from the overlap diagrams that the Laughlin state is a good description of the ground state over a wider region of parameters in the absence of pinning than the quasihole state when there is pinning. To have a feeling of what the ground states look like in different regions of the parameter space, we display site occupations $\langle n_i \rangle$ in Fig. \ref{fig:Site occupations 2p} for three sample points (with fixed $\Omega$) marked by stars in Fig. \ref{fig:Phase diagrams}(a). Left panels of each row in Fig. \ref{fig:Site occupations 2p} show the case of $V=0$ and right panels are for $V\neq 0$; rows are organized from top to bottom in descending order with respect to $\phi$. In the first row, where both Laughlin and quasihole-state overlaps are high and our indicators are very close to the expected values, one recognizes the expected features of the density profiles upon a careful inspection. In panel (a), one can see that the density at the central site is almost $\phi/2$, which is the expected value of the incompressible plateau region that appears for larger number of particles, and there is a characteristic bump close to the cloud edge here lying on the nearest and next-nearest neighbors of the central site due to the smallness of the system. The quasihole can also be seen to be successfully pinned at the central site in panel (b), slightly pushing the particles outwards [one can notice the slight density increase in panel (b) in the outer lattice sites comprising the cloud edge in panel (a)] to let an extra flux quantum in the center of the cloud. Panel (c) of the second row with less flux still shows the same features of the Laughlin state as in panel (a), but now slightly enlarged in size to contain as much flux as necessary, which could also be deduced from the high overlap with the Laughlin state (see Fig. \ref{fig:Overlaps}); however, in panel (d) the single-site pinning potential does not seem to be efficient to localize the quasihole, creating only a small depletion compared to that in panel (b) and the cloud size remains almost the same as in panel (c). Actually, in this case the ground state still has a considerable overlap of $\sim 70\%$ with the Laughlin state \cite{Supplemental}. Also, even if it had been possible to pin the quasihole with a different pinning profile, the expanded cloud would have touched the edges of the system spoiling the fidelity anyway. In the last row with still lower flux, it can be deduced that the incompressible state does not even form as the density profile without pinning is rather Gaussian-like instead of having a more or less flat central region as in panels (a) and (c). To summarize, we observe in Fig. \ref{fig:Site occupations 2p} the ability of the harmonic trap to keep particles in the center of the system, which at the same time enables the system to make an automatic adjustment of the cloud size to enter into the FQH regime depending on the flux. 

%%%
{\it Conclusion.---}
%%%
Creation and observation of lattice FQH states of ultracold atoms is a long-sought-after goal, the first steps of which have been taken in a recent experiment with the creation of a two-particle FQH state. In this work, we claimed that with the addition of a few ingredients to this setup like the harmonic trap and a single-site pinning potential, a quasihole state with very few particles, which can be quite successfully described by the paradigmatic Laughlin-type wave function, can be created and unambiguously observed through a standard density measurement. Making use of the flux adjusting ability of the harmonic trap in an incompressible state, we determined through an extensive parameter search favorable regimes for the creation of the quasihole state. Moreover, we suggested that the mean-square-radius and density depletion measurements taken together would yield quite dependable signatures of this state even with very few particles. We hope that our work will be a useful guide for the upcoming experiments that would clearly reveal this fractionally depleted exotic state in a cold-atom setup.

\begin{acknowledgements}
The author acknowledges financial support through the T\"UB\.{I}TAK-CNR International Bilateral Cooperation Program 2504 (project no.~119N192) and the T\"UBA-GEB\.{I}P Award of the Turkish Academy of Sciences during the early stages of this work. 
\end{acknowledgements}

\end{document}

\title{Supplemental Material \\ for \\``Bulk density signatures of a lattice quasihole with very few particles''}

\maketitle

\section{Mean-square-radii in the absence and presence of pinning}

\begin{figure}
\includegraphics[width=0.45\textwidth]{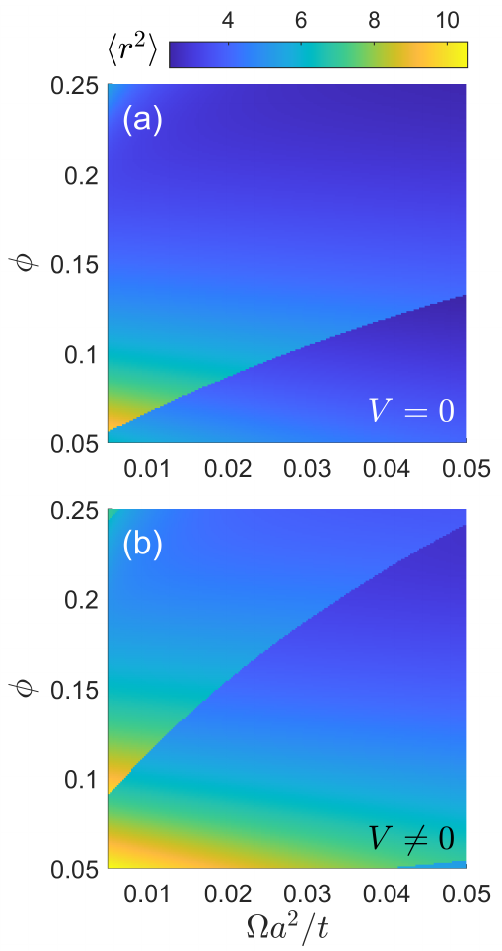}
\caption{Mean-square-radii $\langle r^2\rangle$ for $N=2$ in the plane of the harmonic trap strength ($\Omega$) and flux quanta per unit cell ($\phi$) (a) in the absence of a pinning potential ($V = 0$) and (b) with a pinning potential of strength $V = 10t$.}
\label{fig:Mean square radii}
\end{figure}

Separate results for the mean-square-radii in the absence ($V = 0$) and presence ($V \neq 0$) of pinning, in contrast to the ratio $\mathcal{R} \equiv \langle r^2\rangle_{V = 0}/\langle r^2\rangle_{V \neq 0}$ between them which is considered to be one of the indicators of Laughlin-type physics in the main text, is given in Fig. \ref{fig:Mean square radii}. Although the inverse proportionality of these quantities with the flux quanta per unit cell $\phi$ (in the regions where the overlaps with the ansatz states are high) as expressed by Eq. (5) of the main text is apparent in the figure, they also depend on the harmonic trap strength $\Omega$. To eliminate this effect, we suggested to use the much more robust quantity $\mathcal{R}$ to reflect the Laughlin-type physics. Actually, this quantity is so robust that the details of the density profile are of no practical consequence as revealed by a simple disk model of the densities (see Appendix E of \cite{Umucalilar2018b}). 

\section{Effect of the pinning potential strength}

\begin{figure}
\includegraphics[width=0.42\textwidth]{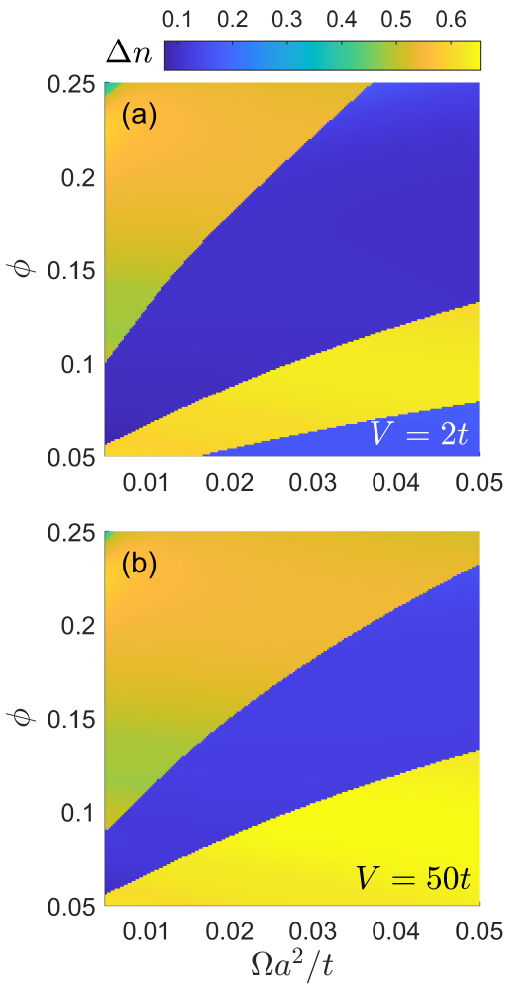}
\caption{Maximum density depletion $\Delta n$ for $N=2$ in the plane of the harmonic trap strength ($\Omega$) and flux quanta per unit cell ($\phi$) with different pinning potential strengths (a) $V = 2t$ and (b) $V = 50t$.}
\label{fig:Different Vs}
\end{figure}

\begin{figure}
\includegraphics[width=0.45\textwidth]{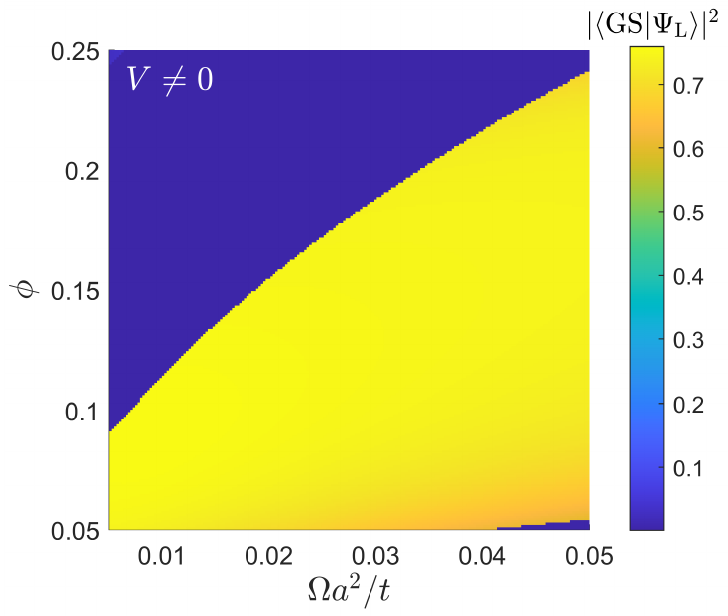}
\caption{Overlap $\mathcal{O}_{\rm L} = |\langle {\rm GS}|\Psi_{\rm L}\rangle|^2$ between the system ground state and the Laughlin state for $N=2$ particles with $V = 10t$. See Fig. 3(b) of the main text for the quasihole overlap for the same parameters.}
\label{fig:Laughlin overlap qh1}
\end{figure}

For a given trap strength $\Omega$ and flux quanta per unit cell $\phi$, the strength of the pinning potential must be above a certain threshold to be able to pin a quasihole. Below this threshold, not enough density is expelled at the position of the pinning potential as can be inferred from Fig. \ref{fig:Different Vs} showing the maximum value of the depleted density in the $\Omega-\phi$ plane. Comparison of panel (a) for the case of $V = 2t$ with panel (b) for $V = 50t$ shows that with increasing strength of the pinning potential the region of agreement with the expected value 0.505 becomes wider. Also, the similarity between panel (b) and the related Fig. 1(b) of the main text which is plotted for an intermediate value of $V = 10t$ implies that there is a saturation value for the pinning strength beyond which the depletion does not change appreciably. Interestingly, in the case of insufficient density depletion and the ensuing vanishing overlap with the quasihole state [see Fig. 3(b) of the main text], the overlap between the ground state of the system and the Laughlin state is considerable ($65-75\%$, see Fig. \ref{fig:Laughlin overlap qh1}), suggesting that the expelled density is mostly redistributed within the Laughlin cloud as in Fig. 4(c-d) of the main text. 

\section{Energy gap}

\begin{figure}
\includegraphics[width=0.42\textwidth]{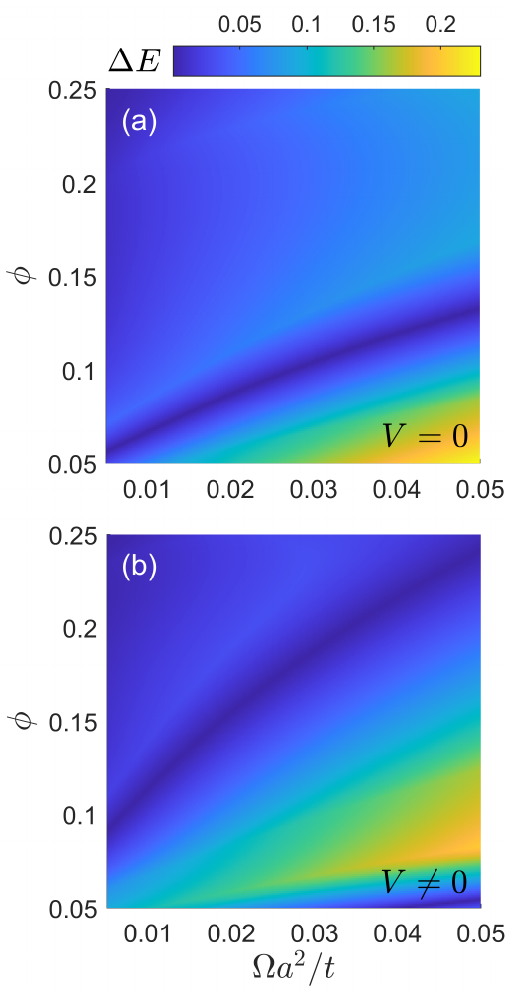}
\caption{Energy gap $\Delta E$ (in units of $t$) for $N=2$ between the ground state and the first excited state of the system in the plane of the harmonic trap strength ($\Omega$) and flux quanta per unit cell ($\phi$) for (a) $V = 0$ and (b) $V = 10t$.}
\label{fig:Energy gaps}
\end{figure}

The overlap diagrams given in Fig. 3 of the main text show clear boundaries with sudden changes. Here, in Fig. \ref{fig:Energy gaps} we display the energy gap above the ground state for the same parameters. We see that the energy gap becomes vanishing following the boundaries of the sudden jumps in the overlaps, which can be related to a dramatic change in the topological character of the system across a boundary.
 
\section{$N = 3$ results}

\begin{figure}
\includegraphics[width=0.45\textwidth]{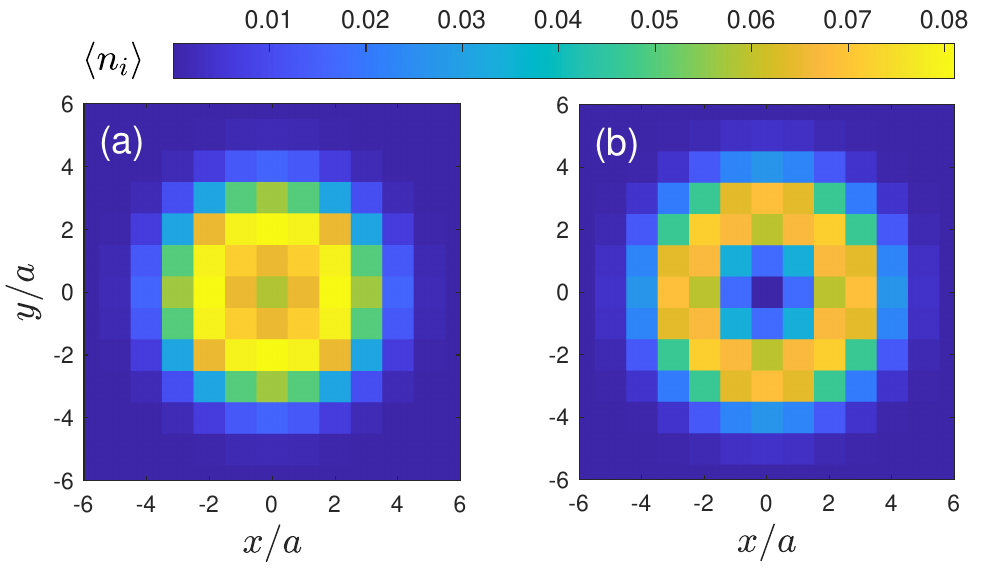}
\caption{Site occupations $\langle n_i\rangle$ for $N=3$ hardcore bosons in a $13\times 13$ lattice calculated for harmonic trap strength $\Omega a^2/t = 0.007$, flux quanta per unit cell $\phi = 0.125$ (a) in the absence of a pinning potential ($V = 0$) and (b) with a pinning potential ($V = 10t$).}
\label{fig:Site occupations 3p}
\end{figure}

In this section, we provide some results for $N = 3$ particles in a $13\times 13$ lattice with hardcore interactions. As the Hilbert space size (790244 states) is large even in the hardcore limit to make a wide parameter scan, we chose to make the calculations for some sample points in the $\Omega-\phi$ plane, which are displayed in Table \ref{table:N3 results}. We observe from the table that whenever the overlaps with the Laughlin state (for $V=0$) and the quasihole state (for $V\neq 0$) are both large, the ratio $\mathcal{R}$ between the mean-square-radii and the maximum density depletion $\Delta n$ are quite close to the ideally expected values [$\mathcal{R}_{\rm ideal} = N/(N+1) = 3/4 = 0.75$ and $(\Delta n)_{\rm ideal} = 0.571$ as obtained from the numerical integration of continuum wave functions]. Site occupations $\langle n_i\rangle$ given in Fig. \ref{fig:Site occupations 3p}, which are calculated for the parameters yielding the largest overlaps $\mathcal{O}_{\rm L,QH}$ in Table \ref{table:N3 results}, also show the expected features of Laughlin and quasihole densities as in Fig. 4(a-b) of the main text.  

\begin{table}
	\caption{Ground-state overlap with the Laughlin state $\mathcal{O}_{\rm L}$ in the absence of a pinning potential $(V = 0)$ and the quasihole state for $\mathcal{O}_{\rm QH}$ with a pinning potential $(V = 10t)$, maximum density of the density depletion $\Delta n$ and the ratio $\mathcal{R}$ between the mean-square-radii, for $N = 3$ particles with various harmonic trap strengths $\Omega$ and flux quanta per unit cell $\phi$. \label{table:N3 results}}
		
	\begin{ruledtabular}
		\begin{tabular}{cccccc}
			$\phi$&$\Omega a^2/t$&$\mathcal{O}_{\rm L} (V = 0)$&$\mathcal{O}_{\rm QH} (V \neq 0)$&$\Delta n$&$\mathcal{R}$\\ 
			\hline
			$ 0.100 $ & $ 0.001 $ & $ 0.9850 $ & $ 0.9680 $ & $ 0.5707 $ & $ 0.7486 $ \\
			$ 0.100 $ & $ 0.007 $ & $ 0.9937 $ & $ 0.0000 $ & $ 0.1540 $ & $ 0.9868 $ \\
			$ 0.100 $ & $ 0.013 $ & $ 0.9863 $ & $ 0.0000 $ & $ 0.1536 $ & $ 0.9871 $ \\
			$ 0.100 $ & $ 0.019 $ & $ 0.9747 $ & $ 0.0000 $ & $ 0.1531 $ & $ 0.9871 $ \\
			$ 0.100 $ & $ 0.025 $ & $ 0.0000 $ & $ 0.0000 $ & $ 0.1508 $ & $ 0.9727 $ \\
			\hline
			$ 0.125 $ & $ 0.001 $ & $ 0.9947 $ & $ 0.9879 $ & $ 0.5854 $ & $ 0.7492 $ \\
			$ 0.125 $ & $ 0.007 $ & $ 0.9960 $ & $ 0.9948 $ & $ 0.5854 $ & $ 0.7506 $ \\
			$ 0.125 $ & $ 0.013 $ & $ 0.9923 $ & $ 0.0000 $ & $ 0.1469 $ & $ 0.9865 $ \\
			$ 0.125 $ & $ 0.019 $ & $ 0.9866 $ & $ 0.0000 $ & $ 0.1454 $ & $ 0.9866 $ \\
			$ 0.125 $ & $ 0.025 $ & $ 0.9792 $ & $ 0.0000 $ & $ 0.1459 $ & $ 0.9865 $ \\
			\hline
			$ 0.150 $ & $ 0.001 $ & $ 0.9918 $ & $ 0.9864 $ & $ 0.5716 $ & $ 0.7496 $ \\
			$ 0.150 $ & $ 0.007 $ & $ 0.9950 $ & $ 0.9944 $ & $ 0.5678 $ & $ 0.7505 $ \\
			$ 0.150 $ & $ 0.013 $ & $ 0.9932 $ & $ 0.9918 $ & $ 0.5643 $ & $ 0.7505 $ \\
			$ 0.150 $ & $ 0.019 $ & $ 0.9901 $ & $ 0.0000 $ & $ 0.1515 $ & $ 0.9856 $ \\
			$ 0.150 $ & $ 0.025 $ & $ 0.9860 $ & $ 0.0000 $ & $ 0.1518 $ & $ 0.9857 $ \\	  
		\end{tabular}
	\end{ruledtabular}
\end{table}